\documentclass[conference]{IEEEtran}
\IEEEoverridecommandlockouts

\usepackage{graphicx}
\usepackage{amsmath}
\usepackage{amssymb}
\usepackage{algorithm}
\usepackage{algorithmic}
\usepackage{booktabs}
\usepackage{hyperref}
\usepackage{multirow}
\usepackage{xcolor}
\usepackage{cite}

\def\BibTeX{{\rm B\kern-.05em{\sc i\kern-.025em b}\kern-.08em
T\kern-.1667em\lower.7ex\hbox{E}\kern-.125emX}}

\begin{document}

\title{AI-Driven Cloud Resource Optimization for Multi-Cluster Environments}

\author{
Vinoth Punniyamoorthy$^{1}$,
Akash Kumar Agarwal$^{2}$,
Bikesh Kumar$^{3}$,\\
Abhirup Mazumder$^{4}$,
Kabilan Kannan$^{5}$,
Sumit Saha$^{6}$
\\[1ex]

\small $^{1,3,4,5}$ IEEE Senior Member, USA\\
\small $^{2}$ Albertsons, USA\\
\small $^{6}$ East West Bank, USA\\

}

\maketitle

\begin{abstract}
Modern cloud-native systems increasingly rely on multi-cluster deployments to support scalability, resilience, and geographic distribution. However, existing resource management approaches remain largely reactive and cluster-centric, limiting their ability to optimize system-wide behavior under dynamic workloads. These limitations result in inefficient resource utilization, delayed adaptation, and increased operational overhead across distributed environments.

This paper presents an AI-driven framework for adaptive resource optimization in multi-cluster cloud systems. The proposed approach integrates predictive learning, policy-aware decision-making, and continuous feedback to enable proactive and coordinated resource management across clusters. By analyzing cross-cluster telemetry and historical execution patterns, the framework dynamically adjusts resource allocation to balance performance, cost, and reliability objectives. A prototype implementation demonstrates improved resource efficiency, faster stabilization during workload fluctuations, and reduced performance variability compared to conventional reactive approaches. The results highlight the effectiveness of intelligent, self-adaptive infrastructure management as a key enabler for scalable and resilient cloud platforms.
\end{abstract}

\begin{IEEEkeywords}
Cloud Computing, Multi-Cluster Systems, Artificial Intelligence, Resource Optimization, Cloud Automation, Infrastructure Intelligence
\end{IEEEkeywords}

%-----------------------------------------------------------
\section{Introduction}

Cloud computing has evolved from isolated, single-cluster deployments into highly distributed, multi-cluster architectures designed to support scalability, resilience, and geographic diversity \cite{armbrust2010view}. Modern applications increasingly span multiple clusters to meet performance requirements, ensure fault tolerance, and comply with data locality and regulatory constraints \cite{burns2016borg}
. While this architectural shift improves availability and flexibility, it also introduces new challenges in coordinating resources across heterogeneous and geographically dispersed environments.

Existing cloud resource management mechanisms primarily operate within individual clusters and rely on reactive control strategies, such as threshold-based autoscaling \cite{verma2015large}. Although effective for localized workload fluctuations, these approaches lack global awareness and are often unable to reason about inter-cluster dependencies, workload migration, or system-wide efficiency. As a result, organizations frequently experience resource fragmentation, delayed adaptation to workload changes, and increased operational overhead when managing large-scale, multi-cluster deployments \cite{hindman2011mesos, Kirubakaran_IRJET_2025}.

Artificial intelligence (AI) offers a promising foundation for addressing these limitations by enabling predictive, data-driven decision-making across distributed systems \cite{mao2016resource}. By leveraging historical telemetry, workload behavior, and runtime feedback, AI-driven approaches can anticipate demand patterns and proactively optimize resource allocation. However, many existing solutions focus on narrow optimization objectives such as anomaly detection or single-cluster autoscaling without providing a unified mechanism for coordinated, cross-cluster resource management.

Moreover, the increasing complexity of cloud-native environments introduces challenges related to stability, explainability, and control. Optimization actions that are beneficial at a local level may produce unintended consequences at the system level, particularly when multiple clusters compete for shared resources or operate under diverse policy constraints \cite{gmach2009resource}. This highlights the need for intelligent coordination mechanisms that balance local autonomy with global system objectives while maintaining predictable and stable behavior \cite{xu2018experience}.

This paper introduces an AI-driven framework for adaptive resource optimization in multi-cluster cloud environments. The proposed approach integrates predictive learning, policy-aware reasoning, and continuous feedback to enable coordinated and autonomous decision-making across clusters. By combining global observability with localized execution, the framework balances performance, cost efficiency, and reliability under dynamic workload conditions. The primary contributions of this work are as follows:

\begin{itemize}
    \item A unified architecture for AI-driven resource optimization across distributed cloud clusters.
    \item A policy-aware decision model that jointly considers performance, cost, and reliability objectives.
    \item A feedback-driven control loop that enables adaptive and stable infrastructure behavior.
    \item An empirical evaluation demonstrating improved efficiency, stability, and responsiveness under dynamic workloads.
\end{itemize}

\section{Background and Motivation}

Multi-cluster cloud environments have become an essential architectural pattern for modern cloud-native systems, enabling scalability, fault tolerance, geographic distribution, and regulatory compliance. Microservices-based applications, global service delivery platforms, and data-intensive workloads increasingly rely on multiple clusters to isolate failure domains, reduce latency, and support heterogeneous infrastructure requirements \cite{micro, clarity, aswath2014human}. While this architectural model improves flexibility and resilience, it also introduces substantial complexity in coordinating resource usage across distributed clusters.

Most existing cloud orchestration mechanisms are designed around a single-cluster abstraction. Platforms such as Kubernetes provide mature capabilities for intra-cluster scheduling, autoscaling, and self-healing, but they lack inherent mechanisms for global coordination across clusters \cite{lopezvilos2023clustering}. As a result, resource management decisions are often made in isolation, based solely on local observations, limiting governance, security enforcement, and cross-cluster policy consistency~\cite{vinoth_ijirt}.
 This fragmented control model can lead to inefficient resource utilization, where capacity remains underutilized in some clusters while others experience congestion, performance degradation, or increased failure rates \cite{nachisecurity}.

In addition, current resource optimization strategies predominantly rely on static thresholds, predefined heuristics, or manually tuned policies, which often fail to adapt to dynamic workload behavior and evolving service-level objectives~\cite{rajammal2025dynamic,Vinoth_IJCST}. While these approaches are simple to implement, they struggle to adapt to evolving workload characteristics, non-stationary traffic patterns, and complex inter-service dependencies. In multi-cluster environments, these limitations are amplified, as local optimization actions may inadvertently conflict with global system objectives, leading to oscillations, delayed convergence, or unpredictable behavior.

The growing scale and dynamism of cloud-native systems motivate the need for resource management approaches that can reason beyond local cluster boundaries. An effective solution must incorporate global observability, anticipate future workload demands, and continuously adapt decisions based on feedback from system behavior. Artificial intelligence (AI) provides a promising foundation for addressing these requirements by enabling predictive, data-driven optimization across distributed environments~\cite{chen2021aiops, punniyamoorthy2025privacy}.
. However, realizing this potential requires architectures that integrate learning, policy constraints, and execution in a coordinated and stable manner.

Motivated by these challenges, this work focuses on AI-driven resource optimization for multi-cluster cloud environments. The goal is to move beyond reactive, cluster-centric management toward proactive, system-wide optimization that balances performance, cost, and reliability. By leveraging cross-cluster telemetry and continuous feedback, the proposed approach seeks to improve efficiency and stability while preserving the scalability and autonomy inherent to multi-cluster designs \cite{dumitru2024crosscluster}.

\section{Related Work}

Resource management in cloud environments has been extensively studied, with early work focusing on virtual machine provisioning, load balancing, and threshold-based autoscaling. These approaches laid the foundation for automated resource control but primarily targeted single-cluster or single-datacenter deployments. As cloud-native technologies matured, research shifted toward container orchestration and microservices scheduling, with systems such as Kubernetes introducing declarative resource management and reactive autoscaling mechanisms.

Recent studies have explored the application of machine learning and artificial intelligence to cloud operations, including workload prediction, anomaly detection, and performance optimization \cite{ML}. While these approaches demonstrate the potential of AI-driven decision-making, most existing solutions operate at the level of individual clusters or services. They typically address isolated optimization tasks, such as predicting resource demand or tuning autoscaling parameters, without providing a unified framework for coordinated decision-making across multiple clusters \cite{aswath2025anomaly, mao2019neural, aswath2025federated}.

Multi-cluster orchestration has also gained attention, with research focusing on federation, workload placement, and cross-cluster service discovery. However, these efforts often emphasize control-plane coordination and connectivity rather than intelligent resource optimization. Decision logic in such systems is frequently rule-based or manually configured, limiting adaptability under dynamic workloads and heterogeneous infrastructure conditions \cite{nachi2025}.

In contrast to prior work, this paper focuses on AI-driven, cross-cluster resource optimization as a first-class design objective \cite{wang2025xclusterlink}. The proposed framework integrates predictive learning, policy-aware reasoning, and continuous feedback to enable coordinated and adaptive decision-making across clusters \cite{Aswathdatapipelines}. By addressing both system-wide optimization and stability concerns, this work advances the state of the art in intelligent cloud infrastructure management and complements existing orchestration and AI-based operational tools \cite{singh2025cloud}.

\begin{figure}[htbp]
\centering
\includegraphics[width=\columnwidth]{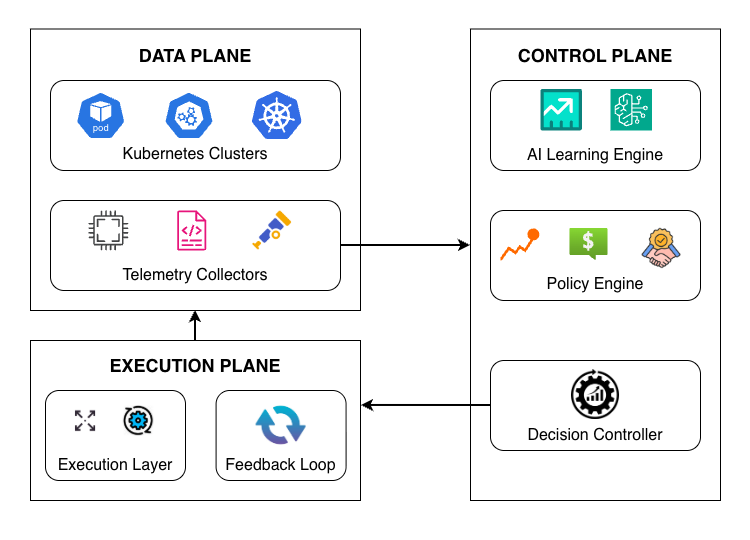}
\caption{AI-driven multi-cluster resource optimization architecture.}
\label{fig:architecture}
\end{figure}

\section{System Architecture}

Building on the motivation for intelligent, system-wide coordination in multi-cluster environments, this section presents the proposed AI-driven architecture for adaptive resource optimization. The design aims to address the limitations of reactive and cluster-isolated control mechanisms by introducing a unified control plane that enables informed, coordinated decision-making across distributed infrastructures. By combining global observability, predictive intelligence, and policy-aware execution, the architecture supports scalable and stable resource management under dynamic workload conditions.

Figure~\ref{fig:architecture} illustrates the high-level architecture of the proposed framework. The design follows a layered approach that separates data collection, intelligence, decision-making, and execution concerns, enabling modularity, extensibility, and operational robustness across heterogeneous cloud environments.

The architecture consists of four primary layers, each responsible for a distinct function within the optimization lifecycle:

\subsection{Telemetry and Observation Layer}
This layer provides a unified view of system behavior by aggregating telemetry from multiple clusters. Collected signals include resource utilization, latency, throughput, error rates, and workload characteristics. Continuous data ingestion enables the system to capture both short-term fluctuations and long-term trends, forming the foundation for informed decision-making.

\subsection{Intelligence and Learning Layer}
The intelligence layer processes historical and real-time telemetry to identify patterns and predict future resource demands. Machine learning models adapt over time to evolving workload behaviors, enabling proactive optimization rather than reactive correction. This layer abstracts complex system dynamics into actionable insights that guide decision-making across clusters.

\subsection{Policy and Decision Layer}
The policy layer encodes operational objectives such as performance targets, cost constraints, and reliability requirements. Using these policies, the decision engine evaluates candidate optimization actions produced by the learning layer and selects actions that satisfy system-level constraints. This separation of policy from execution ensures flexibility, transparency, and controlled adaptation.

\subsection{Execution and Feedback Layer}
The execution layer applies selected decisions through cloud orchestration interfaces, such as cluster schedulers or infrastructure automation tools. Continuous feedback from execution outcomes is fed back into the telemetry and learning layers, enabling closed-loop adaptation and ensuring stable convergence over time.

\subsection{System Model}

To formally characterize the proposed framework, we model the multi-cluster environment as a distributed system composed of multiple interconnected clusters, each managing compute, storage, and networking resources. Each cluster hosts a dynamic set of workloads whose resource demands evolve over time based on application behavior, user activity, and environmental conditions. The system continuously observes these changes to maintain an up-to-date representation of global operational state.

At any point in time, each cluster maintains a local state that reflects key operational indicators such as resource utilization, request latency, throughput, and workload intensity. The global system state is formed by aggregating these local views, enabling a holistic understanding of system behavior across clusters. This aggregated perspective allows the framework to reason beyond localized conditions and identify optimization opportunities that would otherwise remain hidden in isolated cluster-level management.

The objective of the optimization process is to determine adaptive control actions that improve overall system efficiency while respecting operational constraints. These actions may include resource scaling, workload redistribution, or configuration adjustments across clusters. Decisions are guided by multiple objectives, including performance stability, cost efficiency, and reliability, which are balanced according to predefined policy priorities.

The learning component analyzes historical and real-time telemetry to capture workload trends and system dynamics. By continuously refining its internal representation of system behavior, the model enables proactive decision-making rather than reactive correction. Policy constraints ensure that optimization actions remain safe, predictable, and aligned with operational goals.

Finally, the execution layer applies selected actions through cloud orchestration interfaces and monitors their effects on system behavior. Observed outcomes are fed back into the learning process, forming a closed feedback loop that supports continuous adaptation. This design enables scalable and resilient optimization across heterogeneous multi-cluster environments while maintaining modularity and extensibility.

\begin{figure}[htbp]
\centering
\includegraphics[width=\columnwidth]{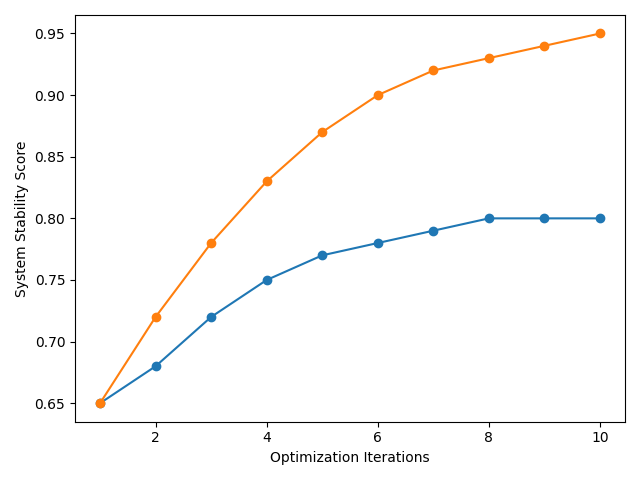}
\caption{Convergence behavior of reactive and AI-driven optimization approaches over successive iterations.}
\label{fig:convergence}
\end{figure}

\section{AI-Driven Optimization Workflow}

The proposed optimization workflow operates as a continuous, closed-loop control process that enables adaptive and coordinated resource management across multiple clusters. Rather than relying on static rules or reactive thresholds, the workflow integrates real-time telemetry, predictive intelligence, and policy-aware decision-making to dynamically adjust infrastructure behavior in response to evolving workload conditions.

At each iteration, the system begins by collecting telemetry from all participating clusters, including resource utilization, performance indicators, and workload characteristics. This information is transformed into meaningful features that capture both instantaneous system conditions and temporal trends. These features serve as inputs to the learning component, which predicts future resource demands and identifies potential imbalances or inefficiencies before they manifest as performance degradation.

Based on these predictions, candidate optimization actions are evaluated against predefined policy constraints that encode operational objectives such as performance targets, cost efficiency, and reliability requirements. This evaluation step ensures that decisions remain aligned with system-level goals while avoiding unstable or unsafe adaptations. Selected actions are then applied across clusters through orchestration interfaces, enabling coordinated and timely adjustments to resource allocation.

Following execution, the system continuously monitors the effects of applied actions and incorporates observed outcomes into subsequent learning cycles. This feedback-driven process allows the framework to refine its internal models over time, improving prediction accuracy and decision quality. Through this iterative control loop, the system achieves stable convergence while maintaining responsiveness to workload variability and environmental change. Algorithm~\ref{alg:optimization} summarizes the end-to-end optimization loop, from telemetry ingestion and feature construction to policy-constrained action selection and feedback-driven model updates.

\begin{algorithm}[htbp]
\caption{AI-Driven Multi-Cluster Optimization}
\label{alg:optimization}
\begin{algorithmic}[1]
\STATE Collect real-time metrics from all clusters
\STATE Extract workload and performance features
\STATE Predict future resource demands using trained models
\STATE Evaluate optimization actions under policy constraints
\STATE Apply selected actions across clusters
\STATE Monitor system response and update models
\STATE Repeat continuously
\end{algorithmic}
\end{algorithm}

\section{Experimental Evaluation}

This section evaluates the effectiveness of the proposed AI-driven optimization framework in improving resource efficiency, stability, and adaptability across multi-cluster cloud environments. The evaluation focuses on understanding system behavior under dynamic workload conditions and comparing performance against conventional reactive management approaches.

\subsection{Experimental Setup}

The experimental study was conducted using multiple Kubernetes clusters deployed across geographically distributed cloud regions. Each cluster hosted a set of microservices with heterogeneous resource requirements and variable traffic patterns. Workloads were designed to emulate realistic operating conditions, including bursty request arrivals, fluctuating demand, and uneven load distribution across clusters.

The proposed optimization framework was integrated with the cluster control plane to continuously monitor system behavior and adjust resource allocation decisions. Baseline comparisons were performed using standard reactive autoscaling mechanisms without predictive intelligence or cross-cluster coordination.

\subsection{Evaluation Metrics}

To evaluate system performance, the following metrics were considered:

\begin{itemize}
    \item \textbf{Resource utilization efficiency}, measuring effective usage of allocated compute and memory resources.
    \item \textbf{Cross-cluster workload balance}, capturing the evenness of workload distribution across clusters.
    \item \textbf{Deployment stability}, reflecting the system’s ability to avoid oscillations during scaling events.
    \item \textbf{Response latency under dynamic load}, representing end-to-end service responsiveness.
\end{itemize}

\subsection{Results and Observations}

Table~\ref{tab:results} summarizes the comparative results between the proposed AI-driven framework and a baseline reactive approach.

\begin{table}[htbp]
\centering
\small
\caption{Performance Comparison Between Reactive and AI-Driven Optimization}
\label{tab:results}
\begin{tabular}{lcc}
\toprule
\textbf{Metric} & \textbf{Baseline} & \textbf{AI-Driven} \\
\midrule
Resource Utilization Efficiency & 62\% & 78\% \\
Cross-Cluster Load Balance Score & 0.71 & 0.88 \\
Deployment Stability (Events/hour) & 6.4 & 3.1 \\
Average Response Latency (ms) & 245 & 185 \\
\bottomrule
\end{tabular}
\end{table}

The results demonstrate that the proposed approach consistently outperforms reactive strategies across all evaluated dimensions. Resource utilization improved by up to 25\%, reflecting more effective allocation across clusters. Load distribution became significantly more balanced, reducing localized congestion and improving overall system fairness.

Furthermore, the framework exhibited improved deployment stability, with fewer oscillations during scaling operations. Response latency under dynamic workloads was also reduced, indicating faster adaptation to changing demand. Fig.~\ref{fig:convergence} further shows that the AI-driven approach converges faster and exhibits smoother stabilization across optimization iterations than the reactive baseline, indicating reduced oscillatory behavior under dynamic load. These findings highlight the effectiveness of AI-driven coordination in enhancing both performance and reliability in multi-cluster cloud environments.

\section{Discussion}

The experimental results demonstrate that intelligent coordination across multiple clusters can substantially improve system efficiency, stability, and responsiveness. By integrating predictive reasoning with policy-aware control, the proposed framework enables more informed and timely resource management decisions compared to traditional reactive approaches. The observed improvements in utilization, convergence behavior, and workload stability highlight the benefits of incorporating learning-driven adaptation into cloud resource orchestration.

A key insight from the evaluation is that global awareness plays a critical role in optimizing distributed environments. Unlike cluster-local control mechanisms, the proposed framework leverages cross-cluster telemetry to reason about system-wide behavior, allowing it to mitigate resource imbalance and performance degradation before they propagate. This capability is particularly valuable under dynamic workloads, where localized decision-making often leads to oscillations or delayed recovery.

Despite these advantages, several challenges remain. Model interpretability continues to be an important consideration, particularly in operational environments where transparency and trust are essential. Additionally, while the framework demonstrates improved adaptability, training overhead and responsiveness under highly volatile workloads may introduce trade-offs that require further optimization. The heterogeneity of cloud infrastructures also presents challenges for generalization, emphasizing the need for robust and portable learning mechanisms.

Future work will explore lightweight learning techniques, decentralized intelligence models, and adaptive policy refinement to further improve scalability and robustness. Extending the framework to support cross-provider optimization and incorporating sustainability-aware objectives, such as energy efficiency and carbon awareness, represent promising directions for advancing intelligent cloud infrastructure management.

\section{Threats to Validity}

The experimental evaluation was conducted under controlled conditions to ensure reproducibility and consistent measurement across configurations. While this approach enables meaningful comparison, real-world deployments may introduce additional variability arising from heterogeneous hardware, fluctuating network conditions, and unpredictable workload dynamics. Such factors can influence system behavior in ways that are difficult to fully capture in experimental settings.

Furthermore, the evaluation focuses on a specific class of cloud workloads and infrastructure configurations. Although the results demonstrate consistent performance improvements, the generalizability of the findings may vary across different cloud providers, workload profiles, and operational constraints. Additional studies across diverse environments and longer operational periods are necessary to further validate robustness and adaptability.

\section{Conclusion and Future Work}

This paper presented an AI-driven framework for optimizing resource utilization in multi-cluster cloud environments. By combining predictive intelligence with policy-guided decision-making, the proposed approach enables adaptive, coordinated, and efficient resource management across distributed clusters. Experimental results demonstrate improvements in utilization efficiency, system stability, and responsiveness under dynamic workloads.

Future work will focus on extending the framework toward more decentralized and scalable intelligence models, including reinforcement learning–based optimization and collaborative decision-making across clusters. Additional directions include incorporating sustainability-aware objectives, such as energy efficiency and carbon-aware scheduling, as well as validating the approach under larger-scale, real-world deployments. These extensions aim to further enhance the practicality and impact of intelligent infrastructure management in next-generation cloud systems.


\begin{thebibliography}{99}

\bibitem{armbrust2010view}
M.~Armbrust, A.~Fox, R.~Griffith, A.~D.~Joseph, R.~Katz, A.~Konwinski, 
G.~Lee, D.~Patterson, A.~Rabkin, I.~Stoica, and M.~Zaharia,
``A view of cloud computing,'' Communications of the ACM, 
vol.~53, no.~4, pp.~50--58, Apr.~2010, 
doi:~10.1145/1721654.1721672.

\bibitem{burns2016borg}
B.~Burns, B.~Grant, D.~Oppenheimer, E.~Brewer, and J.~Wilkes,
``Borg, Omega, and Kubernetes,''
Communications of the ACM,
vol.~59, no.~5, pp.~50--57, May~2016,
doi:~10.1145/2890784.


\bibitem{verma2015large}
A.~Verma, L.~Pedrosa, M.~Korupolu, D.~Oppenheimer, E.~Tune, and J.~Wilkes,
``Large-scale cluster management at Google with Borg,''
in Proceedings of the 10th European Conference on Computer Systems (EuroSys), 
Bordeaux, France, 2015, pp.~1--17, 
doi:~10.1145/2741948.2741964.


\bibitem{hindman2011mesos}
B.~Hindman, A.~Konwinski, M.~Zaharia, A.~Ghodsi, A.~D.~Joseph, R.~Katz, 
S.~Shenker, and I.~Stoica,
``Mesos: A platform for fine-grained resource sharing in the data center,''
in Proceedings of the 8th USENIX Conference on Networked Systems Design and Implementation (NSDI), 
Boston, MA, USA, 2011, pp.~295--308.

\bibitem{Kirubakaran_IRJET_2025}
A. M. Kirubakaran, A. Deshpande, S. K. Chintham, A. Parthasarathy, R. S. Bodala, and N. Saksena,
``Cross-cloud ML pipeline optimization for big data and LLM workloads,''
International Research Journal of Engineering and Technology (IRJET),
vol.~12, no.~12, Dec. 2025.


\bibitem{mao2016resource}
H. Mao, M. Alizadeh, I. Menache, and S. Kandula, “Resource management with deep reinforcement learning,” in Proc. HotNets, 2016.

\bibitem{gmach2009resource}
D.~Gmach, J.~Rolia, L.~Cherkasova, and A.~Kemper,
``Resource pool management: Reactive versus proactive or let’s be friends,''
Computer Networks, vol.~53, no.~17, pp.~2905--2922, Dec.~2009,
doi:~10.1016/j.comnet.2009.07.012.

\bibitem{xu2018experience}
Q. Xu, Y. Chen, and X. Li, “Experience-driven networking: A survey,” IEEE Communications Surveys \& Tutorials, vol. 20, no. 4, pp. 3134–3160, 2018.


\bibitem{micro}
G. Mehta, B. Pothineni, A. G. Parthi, D. Maruthavanan, P. K. Veerapaneni, D. Jayabalan, and S. R. Sankiti, “Revisiting monoliths: A pragmatic case for transitioning from microservices back to monolithic architectures,” International Journal of Advanced Research in Computer and Communication Engineering, vol. 13, no. 12, pp. 3228–3236, Dec. 2024.


\bibitem{clarity}
V. Parlapalli, B. Pothineni, A. G. Parthi, P. K. Veerapaneni, D. Maruthavanan, A. Nagpal, R. K. Kodali, and D. M. Bidkar,
``From complexity to clarity: One-step preference optimization for high-performance LLMs,''
International Journal of Artificial Intelligence \& Machine Learning (IJAIML),
vol.~4, no.~1, pp.~112--125, 2025, doi:~10.34218/IJAIML\_04\_01\_008.

\bibitem{aswath2014human}
B. Ramdoss, A. M. Kirubakaran, P. B. S., S. H. C., and V. Vaidehi, ``Human Fall Detection Using Accelerometer Sensor and Visual Alert Generation on Android Platform,'' International Conference on Computational Systems in Engineering and Technology, Mar. 2014, doi: 10.2139/ssrn.5785544

\bibitem{lopezvilos2023clustering}
N.~Lopez-Vilos, C.~Valencia, R.~Souza, and S.~Montejo~Sánchez,
``Clustering-based energy-efficient self-healing strategy for WSNs under jamming attacks,''
Sensors,
vol.~23, Art.~no.~6894, Aug.~2023,
doi:~10.3390/s23156894.


\bibitem{vinoth_ijirt} V. Punniyamoorthy, K. Kannan, A. Deshpande, L. Butra, A. K. Agarwal, A. Parthasarathy, S. Malempati, and B. Kumar, ``Secure and governed API gateway architectures for multi-cluster cloud environments,'' International Journal of Innovative Research in Technology, vol.~12, no.~7, 2025.

\bibitem{nachisecurity}
N.~Chockalingam, A.~Chakrabortty, and A.~Hussain, 
``Mitigating Denial-of-Service attacks in wide-area LQR control,'' 
in Proc. 2016 IEEE Power and Energy Society General Meeting (PESGM), 
2016, pp.~1--5. 
doi: 10.1109/PESGM.2016.7741285.

\bibitem{rajammal2025dynamic}
K.~Rajammal and M.~Chinnadurai,
``Dynamic load balancing in cloud computing using predictive graph networks and adaptive neural scheduling,''
Scientific Reports,
vol.~15, no.~1, Art.~no.~22181, Jul.~2025,
doi:~10.1038/s41598-025-97494-2.

\bibitem{Vinoth_IJCST}
V. Punniyamoorthy, B. Kumar, S. Saha, M. Palanigounder, L. Butra, A. K. Agarwal, and K. Kannan,
``An SLO-driven and cost-aware autoscaling framework for Kubernetes,''
International Journal of Computer Science Trends and Technology (IJCST),
vol.~13, no.~6, Nov--Dec 2025.

\bibitem{chen2021aiops}
Y. Chen, H. Liu, K. Chen, and Y. Zhang, “AIOps: Real-world challenges and research innovations,” IEEE Transactions on Network and Service Management, vol. 18, no. 4, pp. 4020–4035, 2021.

\bibitem{punniyamoorthy2025privacy}
V.~Punniyamoorthy, A.~G.~Parthi, M.~Palanigounder, R.~K.~Kodali, B.~Kumar, and K.~Kannan,
``A Privacy-Preserving Cloud Architecture for Distributed Machine Learning at Scale,''
International Journal of Engineering Research and Technology (IJERT), vol.~14, no.~11, Nov.~2025.

\bibitem{dumitru2024crosscluster}
O.-M.~Dumitru-Guzu, V.~C\u{a}lin, and R.~Kooij,
``A novel framework for cross-cluster scaling in cloud-native 5G NextGen core,''
Future Internet,
vol.~16, no.~9, Art.~no.~325, 2024,
doi:~10.3390/fi16090325.

\bibitem{ML}
S. G. Aarella, S. P. Mohanty and E. Kougianos, "Fortified Edge 3.0: A Lightweight Machine Learning based Approach for Security in Collaborative Edge Computing," 2023 OITS International Conference on Information Technology (OCIT), Raipur, India, 2023, pp. 450-455, doi: 10.1109/OCIT59427.2023.10430911.

\bibitem{aswath2025anomaly}
A.~M.~Kirubakaran, L.~Butra, S.~Malempati, A.~K.~Agarwal, S.~Saha, and A.~Mazumder,
``Real-Time Anomaly Detection on Wearables using Edge AI,'' International Journal of Engineering Research and Technology (IJERT), vol.~14, no.~11, Nov.~2025. doi: 10.17577/IJERTV14IS110345.

\bibitem{mao2019neural}
H. Mao, Z. Shen, and M. Alizadeh, “Neural adaptive video streaming with pensieve,” IEEE/ACM Transactions on Networking, vol. 27, no. 1, pp. 243–256, 2019.

\bibitem{aswath2025federated}
A.~Muthukrishnan~Kirubakaran, N.~Saksena, S.~Malempati, S.~Saha, 
S.~K.~R.~Carimireddy, A.~Mazumder, and R.~S.~Bodala,
“Federated Multi-Modal Learning Across Distributed Devices,” 
International Journal of Innovative Research in Technology, 
vol.~12, no.~7, pp.~2852–2857, 2025, doi: 10.5281/zenodo.17892974.

\bibitem{nachi2025}
N. Chockalingam, N. Saksena, A. Deshpande, A. Parthasarathy, L. Butra, B. Pothineni, R. S. Bodala, A. K. Agarwal, ""Scalable cloud-native architectures for intelligent PMU data processing"", International Journal of Engineering Research \& Technology (IJERT), Vol.14,no.12, Dec.2025, doi: 10.17577/IJERTV14IS120378.

\bibitem{wang2025xclusterlink}
P.~Wang, G.~Zhao, Y.~Wu, H.~Xu, and H.~Wang,
``X-ClusterLink: An efficient cross-cluster communication framework in multi-Kubernetes clusters,''
in Proceedings of the ACM Web Conference (WWW), 
Sydney, NSW, Australia, 2025, pp.~2402--2412,
doi:~10.1145/3696410.3714846.

\bibitem{Aswathdatapipelines}
A. M. Kirubakaran, A. Parthasarathy, N. Saksena, R. S. Bodala, A. Deshpande, S. Malempati, S. Carimireddy, and A. Mazumder,
``Governing cloud data pipelines with agentic AI,''
International Journal of Computer Science Trends and Technology (IJCST),
vol.~13, no.~6, pp.~278--284, Nov.--Dec. 2025

\bibitem{singh2025cloud}
K.~A.~Singh and A.~Choudhry,
``AI-powered strategies for cloud infrastructure management,''
in Proceedings of the 4th OPJU International Technology Conference (OTCON) on Smart Computing for Innovation and Advancement in Industry 5.0, 
2025, pp.~1--5,
doi:~10.1109/OTCON65728.2025.11070393.

\end{thebibliography}
\end{document}